\documentclass[twocolumn]{svjour3}%
\usepackage{amsfonts}
\usepackage{amsmath}
\usepackage{amssymb}
\usepackage{graphicx}%
\smartqed
\begin{document}
%
\title{Bosonic interactions with a domain wall}%
%

\author{J.R. Morris}%
%

\institute
{Department of Physics, Indiana University Northwest, 3400 Broadway, Gary, IN
46408, USA\\email: jmorris@iun.edu}%
%

\date{Received: date / Accepted: date}%
%

\maketitle
%

\begin{abstract}%

We consider here the interaction of scalar bosons with a topological domain
wall. Not only is there a continuum of scattering states, but there is also an
interesting \textquotedblleft quasi-discretuum\textquotedblright\ of positive
energy bosonic bound states, describing bosons entrapped within the wall's
core. The full spectrum of the scattering and bound state energies and
eigenstates is obtainable from a Schr\"{o}dinger-type of equation with a
P\"{o}schl-Teller potential. We also consider the presence of a boson gas
within the wall and high energy boson emission.%

%

%

\end{abstract}%

\section{Introduction}

A simple planar topological domain wall emerges as a solitonic solution of a
scalar field equation possessing a quartic potential with a broken $Z_{2}$
symmetry (see, e.g., \cite{Vilenkin}-\cite{nts}). In this familiar type of
model the domain wall scalar field $\chi$ interpolates between the
disconnected vacuum states located by $\chi=\pm\eta$, and the energy density
of the wall is localized near the wall's core, where $\chi\rightarrow0$. Other
particles, both fermions and bosons, can interact with the wall in various
ways (see \cite{Vilenkin}-\cite{Rajaraman} for example). In particular, the
scattering of scalar bosons from such a wall has been examined in
\cite{section}. We re-examine this type of scalar field interaction with more
of a focus upon bound states of a complex scalar field $\phi$ interacting with
the domain wall field $\chi$.

\bigskip

Other models of interacting scalar fields have demonstrated how it is possible
for entrapped bosons to stabilize or metastabilize nontopological solitons
formed from domain wall bubbles (see, e.g., \cite{nts},\cite{bub}). However
here, our focus, at least partly, is on the mathematical description of a set
of exact, analytical eigenstates and energy spectra associated with $\phi$
boson scattering states and $\phi$ boson bound states. Using a simple model
for the interacting scalars, and a simple ansatz for the field $\phi$, we
obtain a Schr\"{o}dinger-type of equation with a sech$^{2}$ P\"{o}schl-Teller
potential, which has known solutions. These solutions can be incorporated to
obtain a full description of the $\phi$ scattering and bound states. Certain
values of the potential strength allow for reflectionless scattering. However,
the bound state spectrum is especially interesting in that there is a
discretuum of boson energies associated with motion transverse to the wall,
but for each discrete energy there is a continuum associated with
unconstrained motion parallel to the wall.

\section{Interacting scalar fields}

Our model of interacting scalar fields is basically the same as that presented
in \cite{section}. (See also \cite{bub} for a similar type of model with
quartic self interactions present). The lagrangian for our system of
interacting scalars consisting of a real scalar $\chi$ and a complex scalar
$\phi$ is taken to be%
\begin{equation}
\mathcal{L}=\frac{1}{2}\partial_{\mu}\chi\partial^{\mu}\chi+\partial_{\mu}%
\phi^{\ast}\partial^{\mu}\phi-V(\chi,|\phi|) \label{1}%
\end{equation}

where the potential is%
\begin{equation}
V=\frac{1}{4}\lambda(\chi^{2}-\eta^{2})^{2}+f(\chi^{2}-\eta^{2})\phi^{\ast
}\phi+m^{2}\phi^{\ast}\phi\label{2}%
\end{equation}

The parameters $\lambda$, $f$, $\eta$, and $m$ are all taken to be positive
real-valued constants. The vacuum states of the theory are given by $\chi
=\pm\eta$ and $\phi=0$. The $\chi$ particle and $\phi$ particle masses are
given, respectively, by $m_{\chi}=\sqrt{2\lambda}\ \eta$ and $m_{\phi}=m$. The
field equations for the system that follow from $\mathcal{L}$ are%
\begin{subequations}
\begin{align}
\square\chi+\left[  \lambda(\chi^{2}-\eta^{2})+2f|\phi|^{2}\right]  \chi &
=0\label{3a}\\
\square\phi+\left[  f(\chi^{2}-\eta^{2})+m^{2}\right]  \phi &  =0 \label{3b}%
\end{align}

where $\square=\partial_{t}^{2}-\nabla^{2}$. This system admits the exact
solution set $\phi=0$ and%
\end{subequations}
\begin{equation}%
\begin{array}
[c]{ll}%
\chi(x)=\pm\eta\tanh\bar{x}=\pm\eta\tanh\left(  \Omega x\right)  , & \\
\Omega=\eta\sqrt{\frac{\lambda}{2}} &
\end{array}
\label{4}%
\end{equation}

where we have defined the dimensionless variable $\bar{x}=\Omega x$, where
$\Omega$ is the inverse of the wall's width w. The solution for $\chi$ is that
describing a static domain wall (for, say, $+\eta$) or anti-wall (for, say,
$-\eta$) centered on the $y$-$z$ plane at $x=0$ (see, e.g.,\cite{Vilenkin}%
,\cite{VSbook},\cite{KTbook}). The width w of the wall is%
\begin{equation}
\text{w}=\Omega^{-1}=\frac{1}{\eta}\sqrt{\frac{2}{\lambda}}=2m_{\chi}^{-1}
\label{6}%
\end{equation}

We next consider the behavior of the $\phi$ field in the domain wall
background , as considered in previous works (e.g., \cite{VSbook},
\cite{section}, \cite{bub}). We note that the $\chi$-$\phi$ coupling term is
proportional to $f(\chi^{2}-\eta^{2})=-f\eta^{2}$sech$^{2}\bar{x}$. Using
this, we rewrite the $\phi$ field equation (\ref{3b}) as%
\begin{equation}
\partial_{t}^{2}\phi-\nabla^{2}\phi+m^{2}\phi+U(x)\phi=0 \label{7}%
\end{equation}

with%
\begin{equation}
U(x)=-f\eta^{2}\text{sech}^{2}\bar{x} \label{8}%
\end{equation}

Adopting the plane wave ansatz \cite{section}
\begin{equation}
\phi(x,y,z,t)=\psi(x)\exp\left[  -i(\omega t-\kappa_{y}y-\kappa_{z}z)\right]
\label{8a}%
\end{equation}
with $\omega$, $\kappa_{y}$, $\kappa_{z}$ being real-valued parameters,
(\ref{7}) reduces to%
\begin{equation}
\frac{d^{2}\psi}{dx^{2}}-U(x)\psi+[\omega^{2}-(\kappa^{2}+m^{2})]\psi=0
\label{9}%
\end{equation}

where we define $\kappa^{2}=\kappa_{y}^{2}+\kappa_{z}^{2}$. \ The ansatz
describes a $\phi$ boson freely propagating in the $y$ and $z$ directions, but
subject to a force in the $x$ direction, and having total energy $\omega$. In
the limit $U\rightarrow0$ we have a free boson in three dimensions with the
usual dispersion relation $\omega^{2}=\mathbf{p}^{2}+m^{2}$ where
$\mathbf{p}=(p_{x},p_{y},p_{z})$ is the relativistic momentum and $\omega$ is
the energy. Let us define%
\begin{equation}
K^{2}\equiv\omega^{2}-(\kappa^{2}+m^{2}) \label{10}%
\end{equation}

\ \ The quantity $\sqrt{\kappa^{2}+m^{2}}$ can be considered to be the portion
of the boson energy associated with the two dimensional translational
motion\ in the $y$ and $z$ directions. With (\ref{8}) and (\ref{9}),(\ref{10})
becomes%
\begin{equation}
\frac{d^{2}\psi}{dx^{2}}+\left[  K^{2}+f\eta^{2}\text{sech}^{2}\bar{x}\right]
\psi=0 \label{11}%
\end{equation}

This is just a Schr\"{o}dinger equation for the eigenfunction $\psi$ with a
P\"{o}schl-Teller potential (see, e.g., \cite{Landau}, \cite{Lekner} and
references therein)%
\begin{equation}
U(x)=-\Omega^{2}\nu(\nu+1)\text{sech}^{2}\bar{x} \label{12}%
\end{equation}

where $\nu$ is a real-valued positive constant, not necessarily an integer.
Comparison of (\ref{8}) and (\ref{12}) implies that
\begin{equation}
\nu(\nu+1)=f\eta^{2}/\Omega^{2}=2f/\lambda\label{13}%
\end{equation}
where (\ref{6}) has been used. The depth of the potential well depends upon
$\nu$, or equivalently, $\sqrt{f}\eta/\Omega$.

\bigskip

Exact, analytical sets of scattering eigenstates and bound eigenstates and the
associated spectral values are known, and are elucidated clearly by Lekner in
\cite{Lekner}, whose results we adopt here. In particular, \cite{Lekner}
studies the solutions of the Schr\"{o}dinger equation%
\begin{equation}
\frac{d^{2}\psi}{dx^{2}}+\left[  K^{2}+\Omega^{2}\nu(\nu+1)\text{sech}^{2}%
\bar{x}\right]  \psi=0 \label{14}%
\end{equation}

which coincides with (\ref{11}) using the identification (\ref{13}). Following
\cite{Lekner}, we define an energy parameter $\mathcal{E}=K^{2}/2M$ which
would correspond to the energy of a nonrelativistic particle of mass $M$
described by an eigenfunction $\psi$ of (\ref{14}). We can use the spectral
values of $K^{2}$ to obtain the $\phi$ boson energy spectral values for
$\omega$ by using (\ref{10}). Scattering states for $\psi$, and hence
scattering states for the $\phi$ bosons, form a continuum with $K^{2}>0$, and
bound states of $\psi$ are described by $K^{2}=2M\mathcal{E}<0$, and by the
relation (\ref{10}) will describe the states of $\phi$ bosons trapped in the
domain wall with energy $\omega(\kappa)<\sqrt{\kappa^{2}+m^{2}}$.

\section{Scattering states}

The scattering states are the positive energy states with $K^{2}>0$. By (10)
these are states with $\omega^{2}>m^{2}+\kappa^{2}$ and bosons with this
energy can exist outside the domain wall and are not confined by it, but, in
general, scatter from it. The eigenstates are given by two independent
solutions of (\ref{14}) forming even and odd functions of $x$ and are
described by hypergeometric functions (see, e.g., \cite{Lekner} and references
therein)
\begin{subequations}
\label{15}%
\begin{align}
\psi_{\nu}^{e}(x)  &  =\left(  \cosh\bar{x}\right)  ^{\nu+1}F\left(
\alpha,\beta;\tfrac{1}{2};-\sinh^{2}\bar{x}\right) \label{15a}\\
\psi_{\nu}^{o}(x)  &  =\left(  \cosh\bar{x}\right)  ^{\nu+1}\sinh\bar
{x}\ \times\nonumber\\
&  \times F\left(  \alpha+\tfrac{1}{2},\beta+\tfrac{1}{2};\tfrac{3}{2}%
;-\sinh^{2}\bar{x}\right)  \label{15b}%
\end{align}

where%
\end{subequations}
\begin{equation}%
\begin{array}
[c]{cc}%
\alpha=\frac{1}{2}(\nu+1+i\frac{K}{\Omega}),\medskip & \\
\beta=\frac{1}{2}(\nu+1-i\frac{K}{\Omega}) &
\end{array}
\label{16}%
\end{equation}

and a representation for $F$ is given by%
\begin{equation}%
\begin{array}
[c]{cc}%
F(\alpha,\beta;\gamma;\zeta)=\dfrac{\Gamma(\gamma)}{\Gamma(\alpha)\Gamma
(\beta)}\times\medskip & \\
\times%
{\displaystyle\sum_{n=0}^{\infty}}
\dfrac{\Gamma(\alpha+n)\Gamma(\beta+n)}{\Gamma(\gamma+n)}\dfrac{\zeta^{n}}{n!}
&
\end{array}
\label{17}%
\end{equation}

\bigskip

(We ignore the possible scattering of $\chi$ particles from the $\chi$ wall,
as the reflection coefficient is zero for this case \cite{section}.) Now, the
scattering of $\phi$ bosons from the domain wall becomes reflectionless, for
any energy, for the case that $\nu=$ positive integer \cite{Lekner}. This can
also be seen from the analysis of boson scattering from a domain wall
\cite{section}, where there is a similar $\chi$-$\phi$ interaction, which by
using the same plane wave ansatz, leads to the same Schr\"{o}dinger equation,
provided that we identify the coupling constant $\tilde{\lambda}$ of
\cite{section} with our coupling $f$ \cite{note}. The quoted reflection
coefficient is%
\begin{equation}
\mathcal{R}=\frac{\cos^{2}\theta}{\sinh^{2}\varphi+\cos^{2}\theta} \label{18}%
\end{equation}

where (in our notation, with our parameters) $\varphi=\sqrt{2/\lambda}%
(\pi\kappa/\eta)$ and $\theta=(\pi/2)\sqrt{1+8f/\lambda}$. Using (\ref{13}) we
have $\sqrt{1+8f/\lambda}=\sqrt{4\nu(\nu+1)+1}$. The reflection coefficient
vanishes for $\nu=$ positive integer, where $\cos\theta=0$.

\section{Bound states}

Bound states of the Schr\"{o}dinger equation (\ref{14}) occur when
$K^{2}=2M\mathcal{E}<0$. In this case the energy spectrum for $\mathcal{E}$ is
given by \cite{Lekner}%
\begin{equation}
\mathcal{E}_{n}=-\frac{\Omega^{2}(\nu-n)^{2}}{2M},\ \ n=0,1,2,...,[\nu]
\label{19}%
\end{equation}

where $[\nu]$ is the integer part of $\nu$ ($\nu$ need not be integer). These
negative energy eigenstates can be obtained from (\ref{15}) and (\ref{16}) by
replacing $K$ by $Q=iK$ \cite{Lekner}. The identification $K_{n}%
^{2}=2M\mathcal{E}_{n}$ gives $\omega_{n}^{2}-(\kappa^{2}+m^{2})=-\Omega
^{2}(\nu-n)^{2}$, or
\begin{equation}
\omega_{n}^{2}(\kappa)=-\Omega^{2}(\nu-n)^{2}+(\kappa^{2}+m^{2}) \label{20}%
\end{equation}

The energy spectrum given by the set $\{\omega_{n}(\kappa)\}$ depends upon the
discrete set of integers $n$, as well as upon the continuous parameter
$\kappa\geq0$. We might refer to this as a \textquotedblleft
quasi-discretuum\textquotedblright. We note the following:%
\begin{subequations}
\begin{align}
\omega_{n}^{2}(\kappa)  &  \geq-\Omega^{2}(\nu-n)^{2}+m^{2}\label{21a}\\
&  =\min[\omega_{n}^{2}]\geq0 \label{21b}%
\end{align}

since $\omega_{n}(\kappa)$ is asssumed to be real-valued. (Otherwise the
$\phi$ particles are not stable, as is assumed to be the case.) The above,
however, requires that $\nu-n\leq m/\Omega$, i.e.,%
\end{subequations}
\begin{equation}
\ n\geq\nu-m/\Omega\ \ \text{and\ \ }n\geq0 \label{22}%
\end{equation}

Therefore, there exists a minimal value for the integer $n$ with
$n_{\text{min}}=0$ if $\nu\leq m/\Omega$ and $n_{\text{min}}=$ smallest
integer $\geq\nu-m/\Omega$ for $\nu\geq m/\Omega$ (a \textquotedblleft ceiling
function\textquotedblright\ value).

\bigskip

We now envision a set $\{\omega_{n}(\kappa)\}$ with energies given in
(\ref{20}) with $n=n_{\text{min}},n_{\text{min}}+1,n_{\text{min}}+2,...,[\nu]$
and $\kappa\in\lbrack0,\infty)$. The minimal value of $\omega_{n}(\kappa)$,
i.e., $\min[\omega_{n}(\kappa)]$, increases with increasing $n$, so that the
different $n$ discrete energy levels overlap. Note that for $\nu=$ positive
integer then for $n=\nu$ we have $\min[\omega_{\nu}(\kappa)]=m$, and we have a
barely free $\phi$ boson at rest.

\bigskip

Consider now a gas of $\phi$ bosons at a temperature $T=\beta^{-1}$. The gas
particles collide and interact, and we expect that the ensemble average boson
momentum to be zero, i.e., $\left\langle \mathbf{p}\right\rangle =0$, but we
have $\left\langle \kappa\right\rangle \geq0$. For a gas of $\phi$ bosons with
a large ensemble of particle energies, we now abandon the monochromatic
bosonic beam ansatz of (\ref{8a}) in favor of a quantum statistical
description. We take the chemical potential $\mu$ of the boson gas to be
negligible ($\beta\mu\ll1$), and using boson gas statistics we have that the
number of particles with energy $\omega$ is given by $N(\omega)=(e^{\beta
\omega}-1)^{-1}$. Now, the ideal boson gas can allow sufficiently high energy
particles to escape the wall to $|x|=\infty$ with a minimal energy
$\omega_{\infty}=m$. Therefore gas particles within the wall with average
energies $\omega\geq m$ have the possibility of eventually escaping the wall.
If the gas temperature is $T\ll m$, i.e., $\beta m\gg1$, then the number of
particles $N(\omega\geq m)$ is small and the rate of boson leakage is low.
However, for high temperature $T\gg m$, i.e., $\beta m\ll1$, $N(\omega\geq m)$
is large and the rate of $\phi$ boson leakage from the wall is high. The total
number density of relativistic $\phi$ particles is \cite{KTbook} $n\sim T^{3}%
$, and the total energy density of relativistic particles is \cite{KTbook}
$\rho_{\phi}\sim T^{4}$. So for a planar domain wall inhabited by $\phi$
bosons, we expect escaping bosons to result in a drop of $n$ and $\rho_{\phi}%
$, and consequently a drop in $T$. A drop in temperature $T$ then reduces the
rate of $\phi$ boson emission from the wall.

\bigskip

The model considered here is presented at a classical level, i.e., the domain
wall solution follows from the tree-level potential (\ref{2}) for $\phi=0$.
These tree-level results can be used for an analysis of one-loop quantum
corrections, which will result in a shift of the surface tension of the domain
wall. (See, for example \cite{tension1} and \cite{tension2}).

\section{Summary}

A tree-level model describing two interacting scalar fields has been studied,
wherein one of the scalar fields ($\chi$) exhibits a $Z_{2}$ discrete
symmetry$.$ Consequently, the field equations admit a solution set where the
real field $\chi$ forms a topological planar domain wall of width w
$=\Omega^{-1}$ centered on the $y$-$z$ plane, while the second field ($\phi$)
assumes its vacuum value of $\phi=0$. We then consider excitations of the
$\phi$ field, i.e., $\phi$ boson particles, in the domain wall background.
Upon adopting a modulated plane wave type ansatz with $\phi=\psi(x)\exp\left[
-i(\omega t-\kappa_{y}y-\kappa_{z}z)\right]  $, with $\omega$ and
$\boldsymbol{\kappa}=(\kappa_{y},\kappa_{z})$ being real-valued parameters, a
Schr\"{o}dinger equation with a P\"{o}schl-Teller potential $U(x)=-f\eta^{2}%
$sech$^{2}\Omega x$ is obtained for the eigenfunction $\psi(x)$. The
eigenvalue $K^{2}$ for the Schr\"{o}dinger equation can be written in terms of
boson energy $\omega$ and momentum parameter $\kappa=|\boldsymbol{\kappa}|$.

\bigskip

Solutions describing scattering states of the $\phi$ bosons from the domain
wall are presented. A comparison made with the work in \cite{section} verifies
that the reflection coefficient $\mathcal{R}$ vanishes for positive integers
$\nu$ where we have the identification $\nu(\nu+1)=\Omega^{-2}f\eta
^{2}=2f/\lambda$.

\bigskip

Next, the bound state energy spectrum for $\phi$ bosons trapped within the
domain wall is found. This spectrum is described by a discrete set of
nonnegative integers $n$, along with a set formed by the continuous parameter
$\kappa$. We have therefore referred to this spectrum formed by the set
$\{\omega_{n}(\kappa)\}$ as a \textquotedblleft
quasi-discretuum\textquotedblright, since it is described by a discrete
integer $n$, as well as by a continuous parameter $\kappa$. Since there can be
many $\phi$ bosons in each discrete energy level, then one can consider the
existence of a $\phi$ boson gas at some temperature $T=1/\beta$ trapped within
the domain wall. We suggest, in a way similar to that along the lines of
\cite{bub}, that sufficiently energetic particles with $\omega\geq m$ can
escape the wall. The leakage rate, however, depends upon the gas temperature
$T$, with a low rate of $\phi$ boson emission for $T\ll m$ and a high rate of
emission at $T\gg m$.


\begin{thebibliography}{99}
\bibitem {note}In \cite{section} the interaction between a wall field and a
boson field in the wall background is written in the form $\mathcal{L}%
_{\text{int}}=-\frac{\tilde{\lambda}}{2}\phi^{2}\chi^{2}$, where, in
\cite{section}, $\phi$ represents the wall field and $\chi$ represents the
interacting boson field. With the change of variables $\tilde{\lambda
}\rightarrow f$, $\phi\rightarrow\chi$, and $\frac{1}{2}\chi^{2}%
\rightarrow\phi^{\ast}\phi$, then $\mathcal{L}_{\text{int}}$ of \cite{section}
maps into that of our Eq. (\ref{2}) and the equations for a complex $\chi$ in
\cite{section} maps into those of our Eqs. (\ref{7}) - (\ref{8a}).
\end{thebibliography}
\end{document}